\documentclass[twocolumn,prb, showpacs]{revtex4}
\usepackage{graphicx}
\usepackage{dcolumn}
\usepackage{bm}

\begin{document}


\title{Sodium ion ordering of Na$_{0.77}$CoO$_2$ under competing multi-vacancy cluster, superlattice and domain formation }

\author{F. -T. Huang$^{1,2,3}$}
\author{G. J. Shu$^{1}$}
\author{M. -W. Chu$^{1}$}
\author{Y. K. Kuo$^{4}$}
\author{W. L. Lee$^5$}
\author{H. S. Sheu$^6$}
\author{F. C. Chou$^{1,6}$}
\email{fcchou@ntu.edu.tw}

\affiliation{
$^1$Center for Condensed Matter Sciences, National Taiwan University, Taipei 10617, Taiwan}
\affiliation{
$^2$Taiwan International Graduate Program, Academia Sinica,Taipei 10115,Taiwan}
\affiliation{
$^3$Department of Chemistry,National Taiwan University,Taipei 10617,Taiwan}
\affiliation{
$^4$Department of Physics, National Dong-Hwa University, Hualian 97401,Taiwan}
\affiliation{
$^5$Institute of Physics, Academia Sinica,Taipei 11529,Taiwan}
\affiliation{
$^6$National Synchrotron Radiation Research Center, HsinChu 30076, Taiwan}

\date{\today}

\begin{abstract}
Hexagonal superlattice formed by sodium multi-vacancy cluster ordering in Na$_{0.77}$CoO$_2$ has been proposed based on synchrotron X-ray Laue diffraction study on electrochemically fine-tuned single crystals.  The title compound sits closely to the proposed lower end of the miscibility gap of x $\sim$ 0.77-0.82 phase separated range.  The average sodium vacancy cluster size is estimated to be 4.5 Na vacancies per layer within a large superlattice size of $\sqrt{19}$a$\times$$\sqrt{19}$a$\times$3c.  The exceptionally large Na vacancy cluster size favors large twinned simple hexagonal superlattice of $\sqrt{19}$a , in competition with the smaller di-, tri- and quadri-vacancy clusters formed superlattices of $\sqrt{12}$a and $\sqrt{13}$a.  Competing electronic correlations are revealed by the observed spin glass-like magnetic hysteresis  below $\sim$ 3K and the twin, triple and mono domain transformations during thermal cycling between 273-373K.

\end{abstract}



\maketitle

\section{\label{sec:level1}Introduction\protect\\ }

Na$_x$CoO$_2$ has a rich x-dependent phase diagram which covers antiferromagnetic spin ordering, metal-to-insulator transition, and superconductivity as a function of Na content.\cite{Foo2004}  Many early studies of Na$_x$CoO$_2$ system estimated the Na content of the as-prepared powder or single crystal to be close to rational fractions of $\frac{2}{3}$ or $\frac{3}{4}$ based on its hexagonal structure.\cite{Motohashi2003a, Sales2004, Helme2005, Roger2007}  In addition, many theoretical calculations based on combined density functional theory show many predicted stable phases including these simple fractional fillings.\cite{Zhang2005, Meng2008}  However, not all predicted sodium ordered phases have been found experimentally; moreover, the expected x of simple rational fractions such as $\frac{2}{3}$ and $\frac{3}{4}$ have not been confirmed so far based on our electrochemical de-intercalation technique.\cite{Shu2007, Shu2008, Chou2008}    Further studies on diffusion property of Na$_x$CoO$_2$ system revealed strong correlation between Na orderings and reduced diffusivity at specific Na contents such as x $\sim$ 1/4, 1/3, 1/2, and 0.71.\cite{Shu2007, Shu2008}

Applying high pressure floating-zone crystal growth and the following electrochemical de-intercalation techniques on Na$_x$CoO$_2$ system, we have demonstrated Na content can be controlled in a more delicate manner compared to the traditional topochemical Br$_2$ de-intercalation route,\cite{Chou2004, Shu2007} which led to an interesting finding of Na ion superlattice ordering for x $\sim$ 0.71 and 0.84.\cite{Chou2008}  It is shown that both x $\sim$ 0.71 and 0.84 have simple hexagonal superstructure of $\sqrt{12}$a and $\sqrt{13}$a respectively, which is formed with the ordered Na multi-vacancy clusters at room temperature within each Na layer.\cite{Roger2007, Chou2008, Huang2009}  Most interestingly, it shows x = 0.71 to be the average of $\frac{2}{3}$ and $\frac{3}{4}$ and the most stable and ordered phase above Na half filling, as a consequence of ordered alternating tri-/quadri-vacancy cluster stacking requirement based on its P6$_3$/mmc symmetry.  On the other hand, latest NMR studies provided strong evidence of sodium ion ordering of Na$_{2/3}$CoO$_2$ based on powder sample prepared from high temperature air quenching.\cite{Platova2009}  The controversy must be originated from the different sample preparation route for Na$_{2/3}$CoO$_{2-\delta}$ in the latter, i.e., the additional oxygen vacancies generated from high temperature air quenching must be responsible for its stable existence with a different superlattice as proposed based on NMR results.  Our preliminary investigation to this controversy supports the former scenario from detailed thermogravimetric analysis of oxygen loss.  We find although the room temperature electrochemically de-intercalated Na$_{2/3}$CoO$_2$ shows no sign of Na ordering, Na$_{2/3}$CoO$_{1.98}$ does reveal a $\sqrt{12}$a superlattice after additional high temperature annealing and air quenching, which is in agreement with the superlattice size proposed by Platova \textit{et al.},\cite{Platova2009}  the complete results will be reported separately.  Clearly both sodium and oxygen contents are of equal importance when studying the physical properties of this unique system.  While both can alter Co-layer electronic configuration significantly, our sample preparation method uniquely allows pure sodium content control without complication from oxygen vacancies .

Phase separation phenomenon has been observed in the narrow ranges of $\sim$ 0.77-0.82 and 0.83-0.86,\cite{Shu2009} where detailed phase diagram has been mapped and interpreted as a result of Na ordering within each layer as well as staging along the c-direction.  The superstructure of Na$_{0.77}$CoO$_2$ that sits near the lower end of the 0.77-0.82 miscibility gap has not been reported in detail so far, especially it shows the largest simple hexagonal superlattice size of $\sqrt{19}$a in comparing with the rest compositions of x $\gtrsim$ 0.82 with $\sqrt{13}$a and x = 0.71 with $\sqrt{12}$a.  Combining independently obtained Na content from electron probe microanalysis (EPMA) and Laue diffraction study using synchrotron X-ray source, we propose Na ordering for x $\sim$ 0.77 is composed of average 4.5 vacancies out of 19 filled Na sites per $\sqrt{19}$a superlattice unit, which is the largest multi-vacancy cluster and superlattice size found in the Na$_x$CoO$_2$ system so far.  In addition, spin glass-like magnetic ordering below $\sim$ 3K is revealed by its anisotropic irreversible magnetic behavior.

\section{\label{sec:level1}Experimental\protect\\}

Crystal growth of the original high Na content sample and the additional electrochemical de-intercalation procedures have been reported in detail previously.\cite{Shu2007, Chou2008}  Since we need to construct our superstructure model based on independently obtained Na content and diffraction pattern, the Na content has been carefully characterized to be 0.768$\pm$0.004 using both electron probe microanalysis (EPMA) and its linear c-axis lattice parameter versus x relationship, which was constructed from combined Inductively Coupled Plasma (ICP) and EPMA.\cite{Foo2004, Chou2008}  Unlike the treated low x Na$_x$CoO$_2$ crystals of wavy surface and easily to be hydrated in the air, for x $\geq$ 0.71, on the other hand, large, flat and shinny crystal surface can be prepared easily through cleaving, which significantly reduces the standard deviation from multiple EPMA scans.  The typical sample used on for EPMA study is freshly scanned piece of $\sim$ 1$\times$1 mm$^2$ in size, five 10-points line scans are taken both horizontally and vertically and the atomic percentage is calculated through Co normalization to 1.  We describe the sample as x = 0.77 in the following text.  Laue diffraction picture is obtained using synchrotron X-ray source in NSRRC of Taiwan.  Magnetic susceptibility is measured with Quantum Design MPMS-XL with magnetic fields of 1 Tesla and 0.01 Tesla applied parallel and perpendicular to the ab-plane respectively. Specific heat and resistivity measurement has been performed on single crystal sample and all thermal cycling uses rate lower than 2 K/min that is consistent with the temperature dependent Laue and susceptibility measurement.

\section{\label{sec:level1}Results and Discussions\protect\\ }

\begin{figure}
\begin{center}
\includegraphics[width=3.5in]{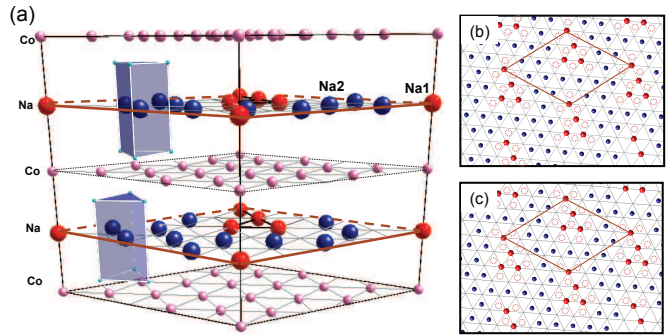}
\end{center}
\caption{\label{fig:fig1}(color online) (a) Crystal structure of Na$_x$CoO$_2$ with only Na and Co ions shown in one $\sqrt{19}$a$\times$$\sqrt{19}$a$\times$c superlattice.  Na1 (red) sits directly on top of Co and oxygen ions are neglected for clarity except one pair of trigonal prismatic oxygen cages (shaded blue) that surround the Na2(blue) in upper and lower Na layers.  (b)-(c) 2D view of upper and lower Na layer, where large vacancy cluster is proposed to be composed of one tri-vacancy plus one di-vacancy in the upper layer, and one quadri-vacancy plus one di-vacancy in the lower layer.  Note Na2 sites rotate 180$^\circ$ between upper and lower Na layers, and (b)-(c) show only one of the two equivalent choices for each di-vacancy that attaches to one corner of the tri-/quadri-vacancy cluster. }
\end{figure}

Na$_x$CoO$_2$ shows a P6$_3$/mmc symmetry, which can simply be described as Na ions filling in between the CoO$_2$ layers of 2D hexagonal network as shown in Fig.~\ref{fig:fig1}.  Although the Na1 site that sits above Co directly is energetically unfavorable, the formation of multi-vacancy cluster would prefer Na-trimer formed using solely Na1 sites, as suggested by Roger $\textit{et al.}$\cite{Roger2007}  Such Na-trimer that corresponds to tri-vacancy or quadri-vacancy clusters  would order in a simple hexagonal superlattice of $\sqrt{12}$a for x $\cong$ 0.71.\cite{Chou2008}

\begin{figure}
\begin{center}
\includegraphics[width=3.5in]{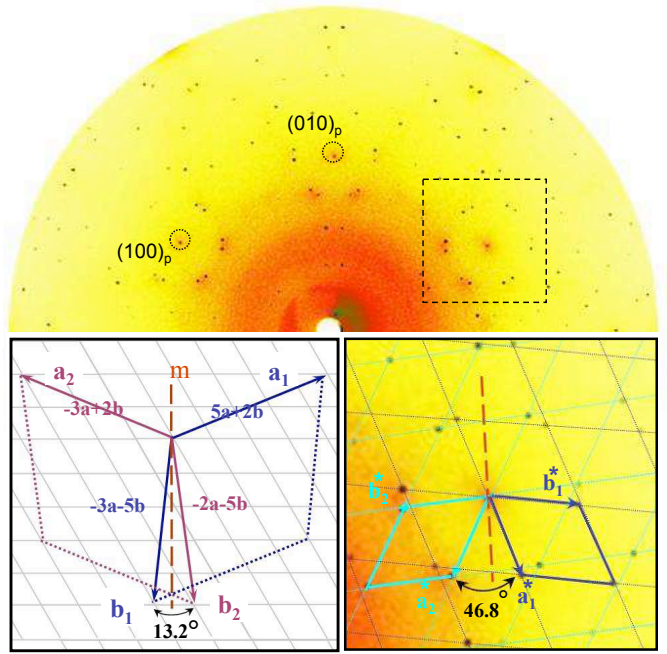}
\end{center}
\caption{\label{fig:fig2}(color online) Laue diffraction pattern of x=0.77.  The 12 spots  ring that surrounds the (001) zone corresponds to two equivalent domains in reciprocal space as shown in the enlarged lower right inset, the reciprocal vector corresponds to $\sqrt{19}$a in the real space. The lower left inset shows the proposed real space model that is constructed by two twinned simple hexagonal superlattice of $\sqrt{19}$a.}
\end{figure}

We are able to solve the 2D superlattice of 0.77 using synchrotron Laue diffraction on single crystal along the (001) direction.  Similar to that found in 0.71 and 0.84,\cite{Chou2008}  The 12-spot ring that is a result of superstructure diffraction and surrounds the low indexing primitive (100)$_p$ planes for 0.77 is shown in the inset of Fig.~\ref{fig:fig2}.   x=0.77 pattern shows two sets of simple hexagonal lattice of a*' $\sim$ $\frac{1}{\sqrt{19}}$ a* in the reciprocal space with an angle of 46.8$^\circ$ in between.  These two sets of a*' in the reciprocal space correspond to two simple hexagonal lattices in real space of $\sqrt{19}$a with an angle of 13.2$^\circ$ in between, which turns out to be two equivalent choices of simple hexagonal superlattices of \textbf{a'} = 5\textbf{a}+2\textbf{b}.  This two domain interpretation has also been verified by electron diffraction from the observed untwinned region (not shown).  The newly found superlattice of $\sqrt{19}$a for x = 0.77 is the largest for x $\gtrsim$ 0.71, where $\sqrt{12}$a superlattice for 0.71 is composed of tri- and quadri-vacancy clusters and $\sqrt{13}$a superlattice for 0.84 is composed of di-vacancy clusters.\cite{Chou2008, Huang2009}

The proposed simple hexagonal superlattice of $\sqrt{19}$a formed by Na vacancies should combine with independently measured Na content and to be able to provide a superstructure model that has a correct stoichiometry of x $\sim$ 0.77.  Since the symmetry of $\gamma$-Na$_x$CoO$_2$ is described by space group P6$_3$/mmc, the Na2 position in adjacent layer is shifted by 180$^\circ$ as shown in Fig.\ref{fig:fig1}, where Na2 sits at the favorable ($\frac{2}{3}, \frac{1}{3}, \frac{1}{4}$) position and Na1 sits at the unfavorable position (0,0,$\frac{1}{4}$) that is directly above Co.  As  Na tri- and quadri-vacancy cluster stacking has been applied successfully to account for the stoichiometry of x $\cong$ 0.71,\cite{Chou2008, Huang2009} i.e., x = 1 - $\frac{1}{2}(\frac{3+4}{12}$) = 0.708, we find similarly the most possible cluster model that accounts for the stoichiometry of x $\sim$ 0.77 with superlattice of $\sqrt{19}$a is the quadri- and penta-vacancy stacking, i.e., there are alternating 4 and 5 Na vacancies per 19 Na within each Na superlattice to form x = 1 - $\frac{1}{2}(\frac{4+5}{19}$) = 0.763.  It is very hard to approach the exact stoichiometry of x=0.763 as described by this ideal superstructure model electrochemically, partly due to the fact that  0.77 sits very close to the 0.77-0.82 miscibility gap boundary, stage ordering of defects would also contribute to the slight deviation from the ideal arrangement.\cite{Shu2009}  In fact, preliminary powder diffraction results indicate the existence of 3c periodicity in x $\sim$ 0.77, which suggests the existence of stage ordering similar to that found in samples of x $\sim$ 0.82-0.86, and the obtained 0.77 is very close to the ideal x=0.763 cluster model plus one extra Na in every $\sqrt{19}$a$\times$$\sqrt{19}$a$\times$3c superlattice unit, i.e., x = 0.763+$\frac{1}{6}\times\frac{1}{19}$ = 0.771.

In order to form quadri- and penta-vacancy clusters in Na layer, the simplest arrangement is to add one more di-vacancy (thus Na1 monomer) attached to the original tri- and quadri-vacancy cluster as described in the x = 0.71 model.\cite{Chou2008, Huang2009}  However, there are six equivalent sites for the n.n. Na1-monomer (di-vacancy) to attach to the Na1-trimer formed with tri- and quadri-vacancy clusters as shown in Fig.~\ref{fig:fig1}, which would introduce additional triangular ambiguity.  It would be more complicated when ordering along c-axis is considered, let alone the potential defect introduced near the ideal model that centered at x=0.763.  The spin glass behavior found below $\sim$3K could be intimately related to the multiple ground states generated by such triangular frustration, which will be discussed in the following section.  The real space structure model shown in Fig.~\ref{fig:fig1} is just one ideal in-phase stacking of identical quadri- and penta-vacancy clusters; potential ordering along c-direction is not discussed here for such a large superstructure size.  We have found the introduced Na multi-vacancy clusters are arranged in an interesting 3$_1$ screw pattern within the triple c-axis for x = 0.71.\cite{Huang2009}  There could exist similar spiral arrangement along c-direction for 0.77 also, but refinement for such large lattice makes it much more difficult.  There have been many studies of Na$_x$CoO$_2$ with nominal stoichiometry near 0.75 that implies a simple fractional filling of $\frac{3}{4}$ before, both from experimental works and theoretical model calculations.\cite{Zhang2005, Julien2008}  However, this assumption has been disproved to be a stable phase with ordered Na ions from diffusion and synchrotron X-ray studies.\cite{Shu2007, Shu2008, Chou2008}  Current observation of stable Na ordered phase near 0.77 instead of $\frac{3}{4}$, in additional to the stable 0.71 instead of $\frac{2}{3}$, is another support to the importance of Na trimer stacking in the P6$_3$/mmc space group symmetry for $\gamma$-Na$_x$CoO$_2$.


\begin{figure}
\begin{center}
\includegraphics[width=3.5in]{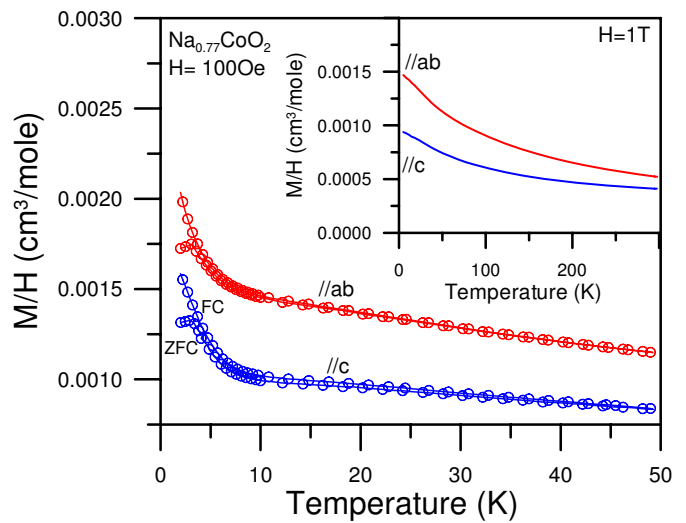}
\end{center}
\caption{\label{fig:fig3}(color online) Magnetic susceptibility measurement results for Na$_{0.77}$CoO$_2$ single crystal under applied field of 100 Oe and 1 Tesla (inset) in both orientations.  Irreversible behavior below $\sim$3K is observed at low field, while high field fitting shows satisfactory Curie-Weiss behavior between 5-300K. }
\end{figure}

Curie-Weiss behavior for x = 0.77 similar to that of x = 0.71 has been found under high field as shown in Fig.~\ref{fig:fig3}.  Based on the reported magnetic susceptibilities of phase separated sample of x = 0.80,\cite{Shu2009} a stoichiometry in the middle of 0.77-0.82 miscibility gap, both A-type antiferromagnetic and spin glass-like transitions are found near 22K and 3K respectively, the former corresponds to the phase of x=0.82 and the latter must be due to the existence of phase with x $\sim$ 0.77.  Curie behavior at high field low temperatures has been found in some of the previous reports for x $\gtrsim$ 0.75,\cite{Luo2004, Prabhakaran2005} and we believe the "Curie-tail" attributed to the paramagnetic impurity could actually have been due to the existence of the x$\sim$0.77 secondary phase while the actual average stoichiometry falls within the miscibility gap between 0.77-0.82.  Magnetic susceptibility for x = 0.77 has been analyzed with Curie-Weiss law fitting ($\chi_{\circ}$ + C/(T-$\Theta$)) satisfactorily, where powder average ($\frac{2}{3}\chi_{ab} + \frac{1}{3}\chi_c$) is taken from high field data shown in Fig.~\ref{fig:fig3}.  Curie constant C = 0.154 $cm^{3} K/ mole$ and Weiss temperature $\Theta \approx$ -124K are obtained from data fitted between 5K $<$ T $<$ 300K satisfactorily.  The derived C implies $\sim$ 41$\%$ of the Co ions behave as localized spin of s=1/2, which is significantly higher than that obtained for x = 0.71 ($\sim$ 11$\%$).\cite{Balicas2008, Chou2008}  However, the unusually high Curie constant cannot be explained as coming from Co$^{4+}$ of low spin state $t_{2g}^{5}e_{g}^{0}$ s= 1/2 only while at most 23$\%$ of total Co ions can be Co$^{4+}$, unless the existence of higher spin state is considered like that found in Li$_x$CoO$_2$.\cite{Hertz2008}  Since the CoO$_6$ octahedral near the large Na vacancy cluster center would be distorted as shown by the significant reduction of Co-O layer thickness as x crossing  from $\sim$0.83 to 0.75,\cite{Huang2004} a cooperative and static Jahn-Teller (JT) distortion which favors intermediate spin state (IS) is possible.\cite{Fauth2001}  The intermediate (IS) and high (HS) spin states for Co$^{4+}$ are described as  $t_{2g}^{4}e_{g}^{1}$ and $t_{2g}^{3}e_{g}^{2}$, the corresponding spins would be s = 3/2 and 5/2 respectively.  Since the expected C's for (LS Co$^{3+}$/IS Co$^{4+}$) and (LS Co$^{3+}$/HS Co$^{4+}$) configurations are 0.431 and 1.006 $cm^3 K/mole$ respectively (assuming g=2), which implies either $\sim$36$\%$ of the carriers are localized at Co$^{4+}$ of IS state or $\sim$15$\%$ in HS state.  The existence of IS or HS states for Co$^{3+}$ is completely ruled out, especially when NaCoO$_2$ that possesses Co$^{3+}$ state shows a clear band insulator-like low spin state only.\cite{Lang2005, Vaulx2005}  Although we can always analyze Curie constant using a localized spin picture, it cannot be an absolute description when metallic behavior is observed and sodium multi-vacancy cluster center is a relatively weak trap for electrons sit near the Co ion directly above and beneath it, let alone the strong electronic correlation which cannot be ignored.\cite{Zhou2005a}  Indeed, NMR study on Na$_{2/3}$CoO$_2$ suggests that the Co$^{3+}$/Co$^{4+}$ scenario is inadequate on describing charge distribution within Co planes.\cite{Alloul2009, Platova2009}   The intriguing carrier distribution reflected on the metallic system of non-trivial Curie contribution for x $\sim$ 0.8  has also be interpreted as partial carrier localization for system with large superlattice formation, where localized charge is bound near the Na vacancy centers while the rest are itinerant within substantially large superlattice.\cite{Chou2008}

Low field susceptibilities shown in Fig.~\ref{fig:fig3} indicates ZFC/FC hysteresis appears below $\sim$ 3K.  Such irreversible ZFC/FC behavior strongly suggests the occurrence of spin glass type phase transition. Frequency dependence study of T$_g$ using AC susceptibility measurement is not conclusive due to instrument resolution and temperature limit.  Similar spin glass-like behavior near 3K has been reported on powder sample of nominal x = 0.75 before.\cite{Takeuchia2002}  Although in-plane magnetic correlation to be FM as suggested by neutron scattering for x $\sim$ 0.8,\cite{Bayrakci2005} the FM is a result of Stoner mechanism among itinerant carriers, which requires narrow band with low energy.\cite{Chou2008, Zhou2005a}  The Stoner mechanism may not survive for x = 0.77 and 0.71 of higher carrier density as suggested by its absence of A-AF ordering.  However the exchange among localized spins of AF character persists as derived from magnetic susceptibility data analysis.   Considering the six equivalent choices of how di-vacancy (Na1-monomer) attaches to the tri- and quadri-vacancy (Na1-trimer) as shown in Fig.~\ref{fig:fig1}, the localized spins at Co$^{4+}$ sites near each vacancy cluster center must suffer from mixed spin states as a result of random CoO$_6$ octahedral distortion, and the strong AF correlation ($\Theta$=-124K) among localized spins can be frustrated based on the triangular arrangement of vacancy clusters.  A-type AF ordering found in the range of 0.82-0.86 is destroyed completely below $\sim$22K once x approaches 0.77 and a spin glass-like behavior emerges; such a magnetic behavior change accompanies a drastic superlattice enlargement and local disordering.  An even higher hole concentration for x = 0.71 is believed to be able to reach a novel quantum spin liquid as a result of Kondo coupling of localized spins on a frustrated lattice.\cite{Chou2008, Lee2005}

\begin{figure}
\begin{center}
\includegraphics[width=3.5in]{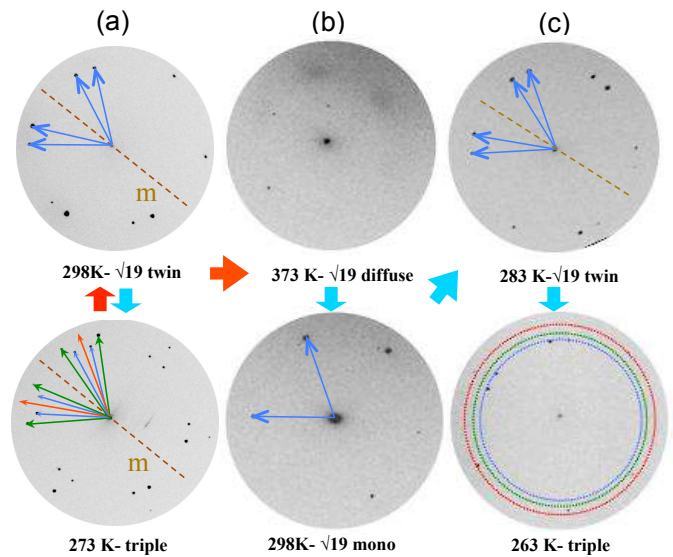}
\end{center}
\caption{\label{fig:fig4}(color online)  Synchrotron X-ray Laue diffraction pictures of Na$_{0.77}$CoO$_2$ along (001) direction under various thermal cycles in sequences of (a) 298-273-298K, (b) 298-373-298K, and (c) 298-283-263-298K.  The heating procedure is shown in red and cooling in thick blue arrows.  The simple hexagonal lattices are indicated by pairs of 60$^\circ$ angle arrows; blue for $\sqrt{19}$a, green for $\sqrt{13}$a and red for $\sqrt{12}$a.  Note the mirror planes are indicated by dashed line to separate two equivalent choices of superlattice, and the circles shown in lower (c) are used to indicate the different diameters for the reciprocal vectors that correspond to the $\sqrt{19}$a (blue), $\sqrt{13}$a (green) and $\sqrt{12}$a (red) in real space.  Final cooling and warming cycles in (c) reinstate the system to the original  $\sqrt{19}$a twin superlattice as shown in upper (a).}
\end{figure}

Temperature dependence of superstructure has been investigated by taking Laue pictures under various temperatures through controlled thermal cycling history.  Fig.\ref{fig:fig4} summarizes the experimental results of selected Laue patterns along (00L) direction taken with synchrotron X-ray.  Three different thermal history cycles are displayed in sequence of (a) 298 - 273 - 298K, (b) 298 - 373- 298K, and (c) 298 - 283 - 263 - 298K.  We note the original room temperature picture shows 12-spot ring that indicates equivalent twin domains of simple hexagonal $\sqrt{19}$a as discussed earlier.  When sample is cooled down to 273K, three twin sets of superstructure which correspond to $\sqrt{19},\sqrt{13}$ and $\sqrt{12}$a are observed, and this transformation is reversible as shown in Fig.~\ref{fig:fig4}(a).   The following heating process toward 373K erases all ordering but a slight diffuse scattering remains, i.e., long range ordering is destroyed but vacancies are trapped locally to maintain partial ordering as indicated by the diffusive dots shown in Fig.~\ref{fig:fig4}(b).  Futher cooling from 373K reorders this $\sqrt{19}$a structure into mono instead of twin domain as shown from its single set of hexagonal diffraction pattern in Fig.~\ref{fig:fig4}(b).  Further cooling of this mono domain sample passes two types of domain transformation as shown in Fig.~\ref{fig:fig4}(c), i.e., the $\sqrt{19}$a twin domain near 283K and then the mixed $\sqrt{19}, \sqrt{13}$ and $\sqrt{12}$a triple superlattices near 263K.  Finally, the superstructure recovers to its original twin domain feature after being warmed up from 263K similar to that shown in upper Fig.~\ref{fig:fig4}(a).  We find there are different ordering mechanisms revealed by the current temperature dependent Laue study, which transform the original twin domain into either mono domain or mixed triple superlattice depending on the thermal history.

The observed triple superlattice of mixed $\sqrt{19}, \sqrt{13}$ and $\sqrt{12}$a below 273K may seem surprising from the point of expected phase separation and phase rule.  However, assuming this layered material to be a pseudo-binary system composed of two components of Na and CoO$_2$, phase rule (under constant pressure) indicates the maximum number of coexisting phases can be three at constant temperature.  The maximized number of phases achieved suggests that entropy plays an important role.  While the confirmed stoichiometry of 0.71 and 0.82-0.86 corresponds to superstructures of $\sqrt{12}$a and $\sqrt{13}$a respectively, the overall average stoichiometry of 0.77 can be conserved under the triple domain superstructure and is close to the ground state.

\begin{figure}
\begin{center}
\includegraphics[width=3.5in]{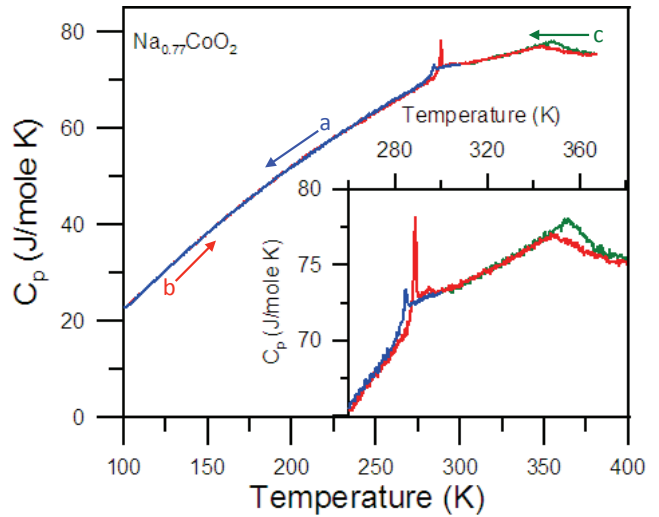}
\end{center}
\caption{\label{fig:fig5}(color online)  Specific heat measurement results of Na$_{0.77}$CoO$_2$ with amplified high temperature range in the inset.  The initial cycle is cooling from room temperature, i.e., in sequence of 300 - 100 - 380 - 300K as indicated by the labels of a (blue), b (red) and c (green). }
\end{figure}

Specific heat measurement results for single crystal sample are shown in Fig.~\ref{fig:fig5}.  Since the Na content has been tuned electrochemically at room temperature, we started the measurement with cooling cycle first in order not to produce any irreversible behavior due to overheating.  The initial cooling passes one less pronounced peak anomaly near 281K, while the following warming cycle shows a sharper spike that has been shifted toward higher temperature near 283K.  Hysteretic character for the crossover near 345-355K is observed.  Using temperature dependent Laue diffraction results shown in Fig.~\ref{fig:fig4} as a comparison, we find that the transition near 281-283K must be related to the twin-domain of $\sqrt{19}$a to triple-superlattice transformation while mono-to-twin domain transformation requires minimum enthalpy change upon cooling from 373K.  As for the 345-355K transition, it can be assigned to the order-disorder crossover, although the persistent diffusive diffraction spot shown at 373K (see Fig.~\ref{fig:fig4}(b)) indicates that the disordering is not completely destroyed but only fluctuating locally near the original multi-vacancy cluster center.  The hysteretic behavior found for both transformations is consistent with its character of irreversible domain conversion and order-disorder transformation.

\begin{figure}
\begin{center}
\includegraphics[width=3.5in]{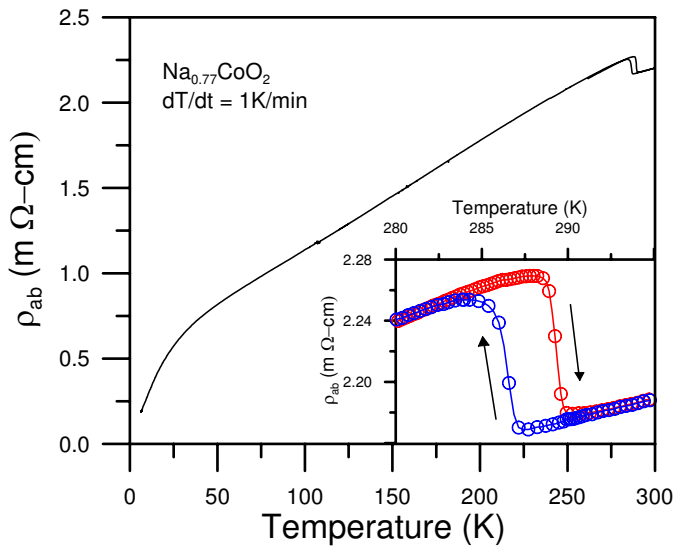}
\end{center}
\caption{\label{fig:fig6}(color online)  Resistivity of Na$_{0.77}$CoO$_2$ in the range of 5-300K.  Thermal hysteresis is found in the range of 284-290K with a significant drop of resistivity at higher temperatures as amplified in the inset.}
\end{figure}

In-plane resistivity of single crystal Na$_{0.77}$CoO$_2$ is shown in Fig.~\ref{fig:fig6}.  We find T-linear dependence between $\sim$ 50-250K, and a significant drop above 290K following a hysteresis between 284-290K.  The 284-290K hysteresis is consistent with the picture of twinned $\sqrt{19}$a superlattice to triple domain crossover as discussed earlier using combined T-dependent Laue diffraction and specific heat measurement results, where different types of charge ordering occur within CoO$_2$ layer as a result of different Na ion ordering patterns.  The slight drop in resistivity for Na ions order in twinned $\sqrt{19}$a superlattice, when compared to that of the triple superlattice state at lower temperature, suggests more scattering near the domain boundaries for the latter.  Following the ideal superstructure model for x=0.77 and the fitted Curie constant discussed ealier, for the 4.5 doped holes per 19 Co ions within $\sqrt{19}$a superlattice unit, 1.6 holes ($\sim$ 36\%) are localized at IS state of Co$^{4+}$ and the remains are itinerant.  The itinerant carriers per Co is only half of that for x = 0.71.  Such reduction is reflected on the resistivity data also, where x=0.77 shows nearly doubled in-plane resistivity comparing with that of x=0.71;\cite{Shu2007} in fact it can be singled out to be the highest among all Na$_x$CoO$_2$ studied so far.\cite{Foo2004}

There have been continue efforts on understanding the intriguing phase transitions found near room temperature.  Charge ordering for x $\sim$ 0.82 at 280K has been suggested by the electronic and lattice response from an optical conductivity study before.\cite{Bernhard2004}  Similar evidence of charge ordering near 280K is hinted for x $\sim$0.8 by both the in-plane resistivity and anharmonic charge vibration perpendicular to the CoO$_2$ layer.\cite{Saint-Paul2008}  Morris $\it{et~al.}$ found superlattice partial meting for x $\geq$ 0.75 from neutron diffraction study, where disordered stripe is suggested just below room temperature.  Weller $\it{et~al.}$ suggests that a 2d melting occurs within Na layers for x = 0.80(1) at 291K based on NMR investigation.\cite{Morris2009, Weller2009}  Although all these previous studies used high temperature prepared samples of nominal sodium content without details of Na ion ordering and convincing chemical analysis, the basic interpretations are confined within the picture of order-disorder transition within Na layers as thermal fluctuation gradually destroys Na ion ordering gradually, which is in agreement with our current conclusion in general.  However, using well characterized single crystal Na$_{0.77}$CoO$_2$, our results provide more details on how do these Na ordered domains re-arrange themselves under the influence of temperature and thermal history, especially the impact of oxygen vacancy is completely ruled out in our experiment.  The detected enthalpy change near 281-283K shown in Fig.~\ref{fig:fig5} suggests that the charge ordering signatures found in resistivity is actually related to the twin domain to triple superlattice phase transformation.  The triple superlattice structure could correspond to the real ground state, and clearly entropy plays an important role on Na ordering under strong electronic correlation.


\section{\label{sec:level1}Conclusions\protect\\ }
In summary, we have found large simple hexagonal superlattice of $\sqrt{19}$a in Na$_{0.77}$CoO$_2$ single crystal using synchrotron X-ray Laue diffraction.  An ideal superlattice model formed with adjacent Na layers of ordered quadri- and penta-vacancies are proposed based on independently obtained Na stoichiometry and model calculation.  The temperature dependent symmetry change revealed by the Laue diffraction is in agreement with a picture of ordering conversions among $\sqrt{19}$a twin domain, $\sqrt{19}$a mono domain, $\sqrt{19}$a +$\sqrt{13}$a+$\sqrt{12}$a triple superlattice, and the vacancy-centered local disordering.  The assignment of these phase transformations are clearly supported by the specific heat and resistivity measurement results.  Current results would help on clarifying many previously published sample dependent and contradicted works based on not fully characterized Na$_x$CoO$_2$ of x $\gtrsim$ 0.7. Na$_{0.77}$CoO$_2$ shows a unique spin glass-like transformation below $\sim$3K and can be explained by its unique di-vacancy to tri-vacancy triangular random attachment within each vacancy cluster.  The newly found low temperature triple superlattice within Na layer suggests that entropy plays an important role in the 2D layered system.  We are contributing more evidence to the existence of stable phase with x near 0.77, instead of a simple fractional filling of $\frac{3}{4}$ which has been commonly suggested before.  The subtle competition and its consequences among cluster size, superlattice size, and domain formation require further investigation so as to better understand the magnetic correlation among the localized spins and the itinerant electrons in a frustrated triangular lattice.

\section*{Acknowledgment}
FCC acknowledges the support from National Science Council of Taiwan under project number NSC-97-3114-M-002 and the helpful discussions with Patrick Lee.



\end{document}